\documentclass[twocolumn,nofootinbib,amsmath,amssymb,aps]{revtex4-1}
\usepackage{bbm} %\mathbbm = doublestruck math for upper, lower, numbers, \mathbbmss=sans serif, \mathbbmtt=typewriter
\usepackage{eufrak} %gothic
\usepackage{mathrsfs} %caligraphic
\usepackage{tcolorbox}
\usepackage{ragged2e}
\usepackage{array,hhline,multirow}
\usepackage{floatrow,subfig,amssymb}
\usepackage[export]{adjustbox}
\usepackage{rotating} % rotate text
\usepackage{mathtools}
\allowdisplaybreaks[2]  %allow page breaks in multi-line eqns, [1]=as little as poss; ... [4]=most permissive
\usepackage{pifont,dsfont}% http://ctan.org/pkg/pifont
\usepackage{bm} %boldmath

\usepackage{graphicx,tikz}
\usepackage[framemethod=TikZ]{mdframed} % fancy boxes
\usepackage{tikz-cd} %for complicated commutative diagrams
\usetikzlibrary{arrows,snakes,shapes.arrows,decorations.markings}
     \tikzset{>=triangle 90}
     \tikzstyle{bbc}=[draw,circle,fill=black,scale=.75]
     \tikzstyle{rc}=[circle,fill=red,scale=.6]
     \tikzstyle{wc}=[draw,circle,scale=.75]

\usepackage[colorlinks]{hyperref}
\hypersetup{
    colorlinks = true,
    linkcolor=blue,
    citecolor=magenta,
    linktocpage}

\def\bar{\overline}

\def\hat{\widehat}

\def\^{\wedge}

\def\cA{{\mathcal A}}
\def\cB{{\mathcal B}}

\def\cC{{\mathcal C}}

\def\cD{{\mathcal D}}

\def\cM{{\mathcal M}}

\def\cN{{\mathcal N}}

\def\cP{{\mathcal P}}

\def\cS{{\mathcal S}}

\def\cT{{\mathcal T}}

\def\cX{{\mathcal X}}

\def\Mm{\cM_{M}[\cT]}

\newcommand{\cZ}{\mathcal{Z}}

\def\Csc{\mathscr{C}}

\def\Z{\mathbb{Z}} 

\def\beq{\begin{equation}}
\def\eeq{\end{equation}}

\newcommand{\bpmat}{\begin{pmatrix}}
\newcommand{\epmat}{\end{pmatrix}}
\newcommand{\bsmat}{\begin{smallmatrix}}
\newcommand{\esmat}{\end{smallmatrix}}
\newfloatcommand{capbtabbox}{table}[][\FBwidth]

\def\bar{\overline}

\def\hat{\widehat}

\def\^{\wedge}

\def\Mm{\cM_{M}[\cT]}

\newcommand{\sC}{\mathsf{C}}
\newcommand{\sM}{\mathsf{M}}

\def\cA{{\mathcal A}}
\def\cB{{\mathcal B}}

\def\cC{{\mathcal C}}

\def\cD{{\mathcal D}}

\def\cM{{\mathcal M}}

\def\cN{{\mathcal N}}

\def\cP{{\mathcal P}}

\def\cS{{\mathcal S}}

\def\cT{{\mathcal T}}

\def\cX{{\mathcal X}}

\def\Mm

\def\Csc{\mathscr{C}}

\def\Z{\mathbb{Z}} 

\def\beq{\begin{equation}}
\def\eeq{\end{equation}}

\usepackage{xcolor}

\newcommand{\sT}{\mathsf{T}}

\def\bM{\begin{matrix}}
\def\eM{\end{matrix}}
\def\bar{\overline}

\def\hat{\widehat}

\def\^{\wedge}

\def\so{\mathfrak{so}}

\def\cC{{\mathcal C}}

\def\cM{{\mathcal M}}

\def\cN{{\mathcal N}}

\def\cP{{\mathcal P}}

\def\cS{{\mathcal S}}

\def\cT{{\mathcal T}}

\def\cX{{\mathcal X}}

\def\Z{\mathbbm{Z}}
\def\ZZ{{\mathbb{Z}}}
\def\cA{{\mathcal{A}}}
\def\cD{{\mathcal{D}}}

\def\p{{\partial}}
\def\eps{{\epsilon}}

\def\no{\nonumber}
\newcommand{\bea}{\begin{eqnarray}}
\newcommand{\eea}{\end{eqnarray}}

\usepackage{comment}
\includecomment{HaveLagrangian}
\includecomment{CiteFusion}

\begin{document}
%TITLE PAGE
\title{Kramers-Wannier-like duality defects in $(3+1)$d gauge theories}

\author{Justin Kaidi}
\affiliation{Simons Center for Geometry and Physics, Stony Brook University, Stony Brook, NY 11794-3636, USA}

\author{Kantaro Ohmori}
\affiliation{Department of Physics, The University of Tokyo, Bunkyo-ku, Tokyo 113-0033, Japan}

\author{Yunqin Zheng}
\affiliation{Kavli Institute for the Physics and Mathematics of the Universe, University of Tokyo, Kashiwa, Chiba 277-8583, Japan}
\affiliation{Institute for Solid State Physics, University of Tokyo, Kashiwa, Chiba 277-8581, Japan}

\begin{abstract}
We introduce a class of non-invertible topological defects in $(3+1)$d gauge theories whose fusion rules are the higher-dimensional analogs of those of the Kramers-Wannier defect in the $(1+1)$d critical Ising model. As in the lower-dimensional case, the presence of such non-invertible defects implies self-duality under a particular gauging of their discrete (higher-form) symmetries.  Examples of theories with such a defect include $SO(3)$ Yang-Mills (YM) at $\theta = \pi$, $\cN= 1$ $SO(3)$ super YM, and $\cN=4$ $SU(2)$  super YM at $\tau = i$. We also introduce an analogous construction in (2+1)d, and give a number of examples in Chern-Simons-matter theories.
\end{abstract}

\maketitle

Symmetries have been a driving force behind modern advances in theoretical physics. Recent developments have led to several extensions of the notion of global symmetry. One such example is higher-form symmetry~\cite{Gaiotto:2014kfa}, which has had numerous applications such as constraining the IR phases of pure Yang-Mills (YM) theory \cite{Gaiotto:2017yup}.

Another type of generalized symmetry is non-invertible symmetry. 
The prototypical example of such a symmetry is the one arising from the Kramers-Wannier self-duality of the (1+1)d Ising model at the critical point. This duality can be implemented by a topological defect line $\cN$ \cite{Verlinde:1988sn,Aasen:2016dop,Freed:2018cec}. If one performs the duality twice, one projects out the $\mathbb{Z}_2$-odd operators, meaning that the composition rule of the topological defect satisfies
\begin{equation}\label{eq.originalKW}
	\cN \times \cN = 1 + \eta_{\mathbb{Z}_2}~,
\end{equation}
with $\eta_{\mathbb{Z}_2}$  the symmetry defect implementing the $\mathbb{Z}_2$ twist of the spin system.
The only topological defects in the Ising CFT are $\cN$ and $\eta_{\mathbb{Z}_2}$, which means that there is no inverse $\cN^{-1}$ such that $\cN^{-1}\times \cN = 1$, and therefore the defect $\cN$ cannot be thought of as implementing a group action. 

The basic idea of non-invertible symmetry is to consider \emph{any} topological defect as a form of generalized symmetry. This means that one must extend the notion of symmetry beyond groups, leading in (1+1)d to a mathematical construction known as a fusion category  \cite{Moore:1988qv,Fuchs:2002cm,etingof2005fusion}.
The Ising category is one of the simplest such fusion categories. 

Though non-invertible symmetries are relatively well-studied in (1+1)d (see e.g.~\cite{Carqueville:2012dk,Bhardwaj:2017xup,Chang:2018iay,Lin:2019hks,Thorngren:2019iar,Komargodski:2020mxz,Huang:2021ytb,Thorngren:2021yso,Huang:2021zvu,Inamura:2021wuo} for recent developments in continuum QFTs), examples in dimensions greater than two remain limited, except in topological QFTs. 
We mention just one example here \cite{Nguyen:2021yld}, in which non-invertible lines were used to study the string tension in (2+1)d $U(1)\rtimes S_N$ semi-abelian gauge theory. 

In this letter, we provide a general procedure for obtaining non-invertible defects in 3+1-dimensions, starting with any theory with 't Hooft anomaly for discrete higher-form symmetries of a particular form.  By gauging a subset of the symmetries appearing in the anomaly, the defect associated with the remaining symmetry becomes non-invertible (such a construction was first suggested in \cite{Tachikawa:2017gyf}). For the particular cases we study, the resulting non-invertible defect will be shown to be a generalization of the Kramers-Wannier defect in (1+1)d. Our defects will have similar implications for self-duality of the gauge theories. Our result can be seen as a continuum analog of the Kramers-Wannier duality of lattice $\mathbb{Z}_2$ gauge theories in (3+1)-dimensions \cite{Freed:2018cec}, whose corresponding defect has been studied in \cite{Koide:2021zxj}.

We illustrate the existence of these defects and their potential dynamical applications through the example of $SO(3)$ gauge theory with zero and one supercharges, as well as $\cN=4$ $SU(2)$  super YM. In the appendices we give details as well as various generalizations of our construction, including to $(2+1)$-dimensional theories (Appendix \ref{app:3d}) and $(3+1)$-dimensional theories with symmetries besides $\ZZ_2$ (Appendix \ref{app:ZNZM}).

\section{General construction}
\label{sec.generalconstruction}

\subsection{Kramers-Wannier-like duality defect}
\label{sec.KW}

Our starting point is a (3+1)d theory $\cT$ with zero-form symmetry $\ZZ_2^{(0)}$ (which can be either linear or anti-linear) and one-form symmetry $\ZZ_2^{(1)}$. Associated to these symmetries are codimension-1 and -2 topological defects.  We will denote the background fields of $\Z_2^{(0)}$ and $\Z_2^{(1)}$ as $A^{(1)}$ and $B^{(2)}$, respectively. The $\Z_2^{(0)}$ symmetry defect inserted on $M_3$ in the presence of $B^{(2)}$ will be denoted by $D(M_3, B^{(2)})$. Throughout we will assume that $M_3$ is oriented.

For simplicity, assume that the spacetime manifold $X_4$ is spin.  The two symmetries can have a mixed 't Hooft anomaly, captured by a 5d integral built from the background gauge fields $A^{(1)}$ and $B^{(2)}$. 
We will be interested in the particular case of a 't Hooft anomaly of the form 
\bea
\label{eq:mainanomaly}
 \pi \int_{X_5} A^{(1)} \cup {\cP (B^{(2)} )\over 2}~,
\eea
with $\cP (B^{(2)} )$ the Pontrjagin square of $B^{(2)}$ and $\partial X_5 = X_4$. The mixed anomaly \eqref{eq:mainanomaly} implies that the $\Z_2^{(0)}$ defect $D(M_3, B^{(2)})$ is anomalous under $\Z_2^{(1)}$ transformations, and hence only the combination 
\begin{eqnarray}\label{Z20Defect}
D(M_3, B^{(2)}) e^{i \pi \int_{M_4} \cP (B^{(2)} )/ 2}
\end{eqnarray}
with $\partial M_4=M_3$ is invariant under gauge transformations of the background field $B^{(2)}$. Note that in \eqref{Z20Defect} the dependence on $M_4$ is only through a term involving the classical background $B^{(2)}$, so \eqref{Z20Defect} should still be regarded as a genuine 3d invertible defect.

We will be interested in understanding the gauging of $\ZZ_2^{(1)}$.
Upon gauging, the background field $B^{(2)}$ is promoted to a dynamical field $b^{(2)}$. From \eqref{Z20Defect}, we see that $D(M_3, b^{(2)})$ is no longer well-defined since it is not invariant under the dynamical gauge transformations of $b^{(2)}$. To make it well-defined, we must either couple to a dynamical bulk, or couple  to a 3d TQFT $\mathfrak{T}(M_3,b^{(2)})$ which cancels the anomaly, thereby absorbing the bulk dependence. Since we will be interested in intrinsically 3d defects, we will pursue the latter strategy. The TQFT cancelling the anomaly is not unique. However, it was shown in \cite{Hsin:2018vcg} that any such TQFT can be factorized into the decoupled tensor product of two theories $\mathfrak{T}(M_3,b^{(2)})= \widehat{\mathfrak{T}}(M_3)\otimes \cA^{2,1}(M_3,b^{(2)})$, where $\widehat{\mathfrak{T}}(M_3)$ does not couple to the dynamical field $b^{(2)}$.  The theory $\cA^{2,1}(M_3,b^{(2)})$ is simply $U(1)_2$ Chern Simons theory, i.e. the minimal TQFT that lives on the boundary of $e^{i \pi \int_{M_4} \cP (b^{(2)} )/ 2}$. Since tensoring a decoupled TQFT only changes the overall normalization of the defect, we choose the TQFT to be the minimal $\cA^{2,1}$ for simplicity. Hence we find a well-defined genuinely 3d defect 
\begin{eqnarray}\label{cN}
\cN(M_3) := D(M_3, b^{(2)}) \cA^{2,1}(M_3,b^{(2)})~.
\end{eqnarray}
Note that $\cN(M_3)$ explicitly depends on the dynamical field $b^{(2)}$.
The defect $\cN$, when regarded as an operator, is linear (resp. anti-linear) if and only if $\Z_2^{(0)}$ before gauging is linear (resp. anti-linear).

We now show that $\cN$ satisfies Kramers-Wannier-like fusion rules; in particular, it is non-invertible.  To begin, consider the case in which $\cN$ is linear. In this case the dual $\bar{\cN}$ is just equal to $\cN$ itself. This follows since $D = \bar{D}$ from $D^2 = 1 $ and $\bar{\mathcal{A}}^{2,1} = \mathcal{A}^{2,-1} \cong \mathcal{A}^{2,1}$, with the last equality following from the fact that $U(1)_{2}$ is time-reversal symmetric as a spin-TQFT \cite{Fidkowski:2013jua,Seiberg:2016rsg}.
We also note that the tensor product theory $\mathcal{A}^{2,1}\otimes \mathcal{A}^{2,-1}$, often called the ``double-semion" theory \cite{freedman2004class,Levin:2004mi}, is equivalent to $\mathbb{Z}_2$ Dijkgraaf-Witten (DW) theory \cite{Dijkgraaf:1989pz} with non-trivial DW twist. The DW twist can be written as $(-1)^{\int_{M_3} a^3}$ where $a \in H^1(M_3,\mathbb{Z}_2)$ is the $\mathbb{Z}_2$ gauge field. Indeed, the $K$-matrix for the double-semion theory, $K = \mathrm{diag}(2,-2)$, can be rotated to that of the $BF$ representation of the DW theory, $K=\begin{pmatrix} 0 & 2 \\ 2 & 1\end{pmatrix}$, by an $\mathrm{SL}(2,\mathbb{Z})$ transformation. 

 These considerations motivate the following result: upon fusion of two $\cN(M_3)$ defects, one obtains a non-trivial $\ZZ_2$ DW theory living on $M_3$. Poincar{\'e} duality allows one to exchange the sum over $\ZZ_2$ 1-cocycles $a$ with a sum over 2-cycles $\Sigma$, upon which the DW twist $e^{i \pi \oint_{M_3}a^3 }$ becomes  the triple intersection number $Q(\Sigma)$ of $\Sigma$ in $M_3$.\footnote{By Banchoff's theorem \cite{banchoff1974triple,banchoff1974triple2}, $Q(\Sigma)$ is equivalent mod 2 to the Euler character of $\Sigma$, i.e. $(-1)^{Q(\Sigma)} = (-1)^{\chi(\Sigma)}$. Hence it is independent of $M_3$ and nontrivial only when $\Sigma$ is unorientable.} This gives the following fusions rules,
\begin{equation}\label{eq:fusionrules}
	\begin{split}
	&\cN(M_3) \times \cN(M_3) \\
		&\quad=\frac{1}{|H^0(M_3, \Z_2)|} \sum_{\Sigma \in H_2(M_3, \ZZ_2)} (-1)^{Q(\Sigma)} L(\Sigma),
	\end{split}
\end{equation}
where $L(\Sigma):= e^{i \pi \oint_{\Sigma}b^{(2)}}$. The normalization factor is related to the volume of the gauge group of the DW theory. A more explicit derivation of both the normalization and the fusion rules will be given in Appendices \ref{app:Lagrangian} and \ref{app:relcoho}. Since this is a sum of more than one operator, we see that $\cN$ is a non-invertible defect.

On the other hand, in the case when $\cN$ is anti-linear, we have $\cN\times \cN= D\cA^{2,1}\times D\cA^{2,1}= D^2 \cA^{2,-1}\times \cA^{2,1}$, since $D$ flips the orientation of $\cA^{2,1}$ as the former passes though the latter. 
Combining this with $D^2=1$, we see that $\cN\times \cN$ again hosts the double-semion theory, giving the same fusion rules as in \eqref{eq:fusionrules}. In summary,  $\cN$ satisfies the fusion rule \eqref{eq:fusionrules} regardless of whether it is linear or anti-linear. 

We may now consider the fusion of $\cN(M_3)$ and $L(\Sigma)$ for some $\Sigma$ embedded in $M_3$.  Note that in \eqref{cN} the one-form symmetry of $\cA^{2,1}$ is coupled to the bulk dynamical field $b^{(2)}$. This means that the Wilson line of $\cA^{2,1}$, which has a non-trivial one-form charge, has to be bounded by the bulk global one-form symmetry generator $L$. In other words, $L$ can end on $\cN$ without costing energy, and furthermore it can be absorbed.
In the process of absorption, the boundary of $L$, identified with the Wilson line, sweeps out a surface $\Sigma$ in $M_3$, producing a sign $(-1)^{Q(\Sigma)}$ from the framing anomaly of the Wilson line. This effect is derived in Appendix \ref{app:Lagrangian}. Hence we obtain the fusion rules
\begin{eqnarray}\label{eq.NLfusionrule}
\cN(M_3)\times L(\Sigma)= (-1)^{Q(\Sigma)} \cN(M_3)~,
\end{eqnarray}
where $\Sigma$ is embedded in $M_3$. 

Finally, the $L\times L$ fusion rule is obvious,
\begin{eqnarray}\label{eq.LLfusionrule}
L(\Sigma)\times L(\Sigma)=1~.
\end{eqnarray}
The fusion rules \eqref{eq:fusionrules}, \eqref{eq.NLfusionrule} and \eqref{eq.LLfusionrule} are reminiscent of the fusion rules of the Ising fusion category in $2$d. For this reason, we refer to the non-invertible defect $\cN$ in the 4d theory $\cT/\Z_2^{(1)}$  as a ``Kramers-Wannier-like defect."  Though we do not work it out here, we expect these fusion rules to yield a fusion 3-category \cite{douglas2018fusion,johnson2020classification}. As we now explain, $\cN$ implements a self-duality transformation on $\cT/\Z_2^{(1)}$.

\subsection{Self-duality}

We now explain why the gauged theory $\cT/ \ZZ_2^{(1)}$  has a notion of self-duality. 
We begin by considering the partition function of $\cT/\ZZ_2^{(1)}$,\footnote{We take the integration measure to be defined with implicit division by the volume of the gauge group. }
\bea
Z_{\cT/ \ZZ_2^{(1)}}[C^{(2)}] = \int \cD b^{(2)} \, Z_{\cT}[b^{(2)}] e^{i \pi \int_{X_4}  C^{(2)}  b^{(2)}}~,
\eea
where $C^{(2)}$ is the  background field for the quantum $\hat\Z_2^{(1)}$ symmetry, whose corresponding defect is $L$.
If we further gauge $C^{(2)}$, the last factor becomes a delta functional for $b^{(2)}$ and we reobtain the original theory $\mathcal{T}$.

To find self-duality in $\cT/\mathbb{Z}_2^{(1)}$, we first include a Dijkgraaf-Witten term $\frac{\pi}{2}\cP(C^{(2)})$ and then gauge. This gives 
\bea\label{selfdual}
&\vphantom{.}&\int \cD c^{(2)}  Z_{\cT/ \ZZ_2^{(1)}}[c^{(2)}] \,\,e^{{i} \int_{X_4}  {\pi\over 2} \cP(c^{(2)})+  \pi  c^{(2)}  A^{(2)}}\no\\
&\vphantom{.}&\quad=\int \cD c^{(2)}\,\cD b^{(2)} \, Z_{\cT}[b^{(2)}] \,e^{i  \int_{X_4} \pi c^{(2)} b^{(2)}+{\pi\over 2}  \cP(c^{(2)})+ \pi  c^{(2)}  A^{(2)}}\no\\
&\vphantom{.}&\quad=\int \cD b^{(2)} \, Z_{\cT}[b^{(2)}] \,\,e^{- i {\pi\over 2} \int_{X_4} \cP(b^{(2)}+ A^{(2)})}
\eea
where we have made a change of variables $c^{(2)} \rightarrow c^{(2)} - b^{(2)} - A^{(2)}$ and dropped a contribution from the TQFT $\int \cD c^{(2)} \, e^{i {\pi \over 2} \int \cP(c^{(2)})}$, which can be continuously deformed to the trivial theory.  
We next use the following anomalous transformation law of $\cT$ under \emph{global} $\Z_2^{(0)}$ transformations:
\begin{eqnarray}\label{sym}
Z_{\cT}[b^{(2)}] e^{-i \frac{\pi}{2}\int_{X_4}\cP(b^{(2)})}= 
\begin{cases}
Z_{\cT}[b^{(2)}], & \Z_2^{(0)} \text{is linear}\\
Z_{\cT}^*[b^{(2)}], & \Z_2^{(0)} \text{is anti-linear}.
\end{cases}
\end{eqnarray}
Eq.\eqref{selfdual} then reduces to
\begin{eqnarray}
Z_{\cT/ \ZZ_2^{(1)}}[A^{(2)} ] \,e^{-i \frac{\pi}{2} \int_{X_4} \cP(A^{(2)})}
\end{eqnarray}
if $\ZZ_2^{(0)}$ is linear, and 
\begin{eqnarray}
Z_{\cT/ \ZZ_2^{(1)}}^*[A^{(2)} ] \,e^{i \frac{\pi}{2} \int_{X_4} \cP(A^{(2)})}
\end{eqnarray}
if $\ZZ_2^{(0)}$ is anti-linear. To complete the self-duality, we need only add a compensating counterterm (and in the anti-linear case, do a complex conjugation $K$).

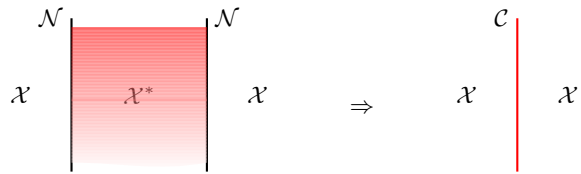
\begin{figure}[!tbp]
\begin{center}
\[\begin{tikzpicture}[baseline=19,scale=0.6]
\draw[  thick] (0,-0.2)--(0,3.2);
\draw[  thick] (3,-0.2)--(3,3.2);
  \node[left] at (-0.7,1.5) {$\cX$};
    \node at (1.5,1.5) {$\cX^*$};
    \node[right] at (3.7,1.5) {$\cX$};
      \node[left] at (0,3.2) {$\cN$};
          \node[right] at (3,3.2) {$ \cN$};  
          \shade[line width=2pt, top color=red,opacity=0.4] 
    (0,0) to [out=90, in=-90]  (0,3)
    to [out=0,in=180] (3,3)
    to [out = -90, in =90] (3,0)
    to [out=190, in =0]  (0,0);
\end{tikzpicture}
\hspace{0.35 in}\Rightarrow\hspace{0.35 in}
\begin{tikzpicture}[baseline=19,scale=0.6]
\draw[red, thick] (0,-0.2)--(0,3.2);
  \node[left] at (-0.7,1.5) {$\cX$};
    \node[right] at (+0.7,1.5) {$\cX$};
      \node[left] at (0,3.2) {$\cC$};
\end{tikzpicture}
\]
\caption{Any theory $\cX$ with a self-duality admits a non-invertible defect with Kramers-Wannier-type fusion rules. $\cX^*$ is the $TST$ (or $TKST$) transform of $\cX$. }
\label{fig:KWfusion}
\end{center}
\end{figure}

In terms of the $S$ and $T$ transformations defined in \cite{Gaiotto:2014kfa} (see also \cite{Bhardwaj:2020ymp, Lee:2021crt}), we conclude that $\cT/ \ZZ_2^{(1)}$ is self-dual under $TST$ if $\ZZ_2^{(0)}$ is linear, and under $TKST$ if $\ZZ_2^{(0)}$ is anti-linear.\footnote{Note that we could have instead started with the theory $\cT' := T \left( \cT/\ZZ_2^{(1)}\right)$, which would be self-dual under $S$ or $KS$. This is the more immediate analog of (1+1)d Kramers-Wannier duality, c.f. Appendix \ref{app:2dcase}, and is the situation studied in \cite{othergroup}. In order to match the fusion rules in \cite{othergroup}, it is necessary to note that upon applying $T$ the one-form symmetry defect is modified to ${L}'(\Sigma):=(-1)^{Q(\Sigma)}L(\Sigma)$. }

We have seen that $\cT/ \ZZ_2^{(1)}$ has a self-duality, as well as a non-invertible defect with Kramers-Wannier-like fusion rules. We now argue that the two facts are related, following \cite{Gaiotto:2019xmp}. First, it is simple to argue that the existence of a self-duality should imply a non-invertible defect of Kramers-Wannier type. Indeed, note that  gauging, and hence the full operation $TST$ (or $TKST$),  can be implemented by a codimension-1 topological defect.
We will denote the total defect by $ \cN(X_3)$.  By stacking two copies of  $\cN(X_3)$, we are left with a condensate as in Figure \ref{fig:KWfusion}. Thus we obtain fusion rules 
\bea
\cN(M_3) \times \cN(M_3) =  \cC(M_3)~.
\eea
The condensate $\cC(M_3)$ can be understood by taking the one-form gauge theory to live in a small tubular neighborhood of $M_3$ with Dirichlet boundary conditions. As shown in Appendix \ref{app:relcoho}, this can be reduced to a zero-form gauge theory living on $M_3$ itself, with the condensate taking the form
\begin{equation}
\label{eq:condop}
\cC(M_3)={1\over |H^0(M_3, \ZZ_2)|} \sum_{\Sigma \in H_2(M_3, \ZZ_2)}(-1)^{Q(\Sigma)} L(\Sigma)~.
\end{equation}
This reproduces the fusion rules of (\ref{eq:fusionrules}). Note that the factor of $(-1)^{Q(\Sigma)}$ descends from the $T$ operation before $S$ in the self-duality, as derived explicitly in Appendix \ref{app:relcoho}. It is easily verified that $\cC(M_3)$ squares to itself up to normalization.

\begin{figure}[!tbp]
\begin{center}
\[\begin{tikzpicture}[baseline=0,scale=0.35]
\draw[red, thick] (-2,2.5)--(-1,1);
\draw[red, thick] (-1,1)--(0,0);
\draw[red, thick] (-2,-2.5)--(-1,-1);
\draw[red, thick] (-1,-1)--(0,0);
\draw[red, thick] (-1,1)--(-1,-1);
\draw[red, thick] (0,0)--(2.5,0);
  \node[left] at (-2,3) {$\cC$};
    \node[left] at (-2,-3) {$\cC$};
     \node[right] at (3,0) {$\cC$};
   \end{tikzpicture}
\hspace{0.35 in}\Rightarrow\hspace{0.2 in}
\begin{tikzpicture}[baseline=0,scale=0.45]
  \draw[thick,black] (-2.2,2.7) to[out=-45,in=180] (2.5,0.2);
   \draw[thick,black] (-2.2,-2.7) to[out=45,in=180] (2.5,-0.2);
    \draw[thick,black] (-2.2,2.3) to[out=-45,in=45](-2.2,-2.3);
    
 \shade[line width=2pt, top color=red,opacity=0.4] 
   (-2.2,2.7) to[out=-45,in=180] (2.5,0.2)
  to[out=-90,in=90] (2.5,-0.2)
 to[out=180, in=45] (-2.2,-2.7)
  to[out=90, in=-90] (-2.2,-2.3)
  to[out=45,in=-45](-2.2,2.3)
   to[out=90,in=-90]  (-2.2,2.7);
\draw[fill = white,thick]  (-0.35,0) circle (15pt);
\node[right] at (0.2,0){$\cN$};
\node[left] at (-2.2,2.7){$\cN$};
\node[left] at (-2.2,-2.7){$\cN$};
\node[right] at (2.5,0.1){$\cN$};
 \end{tikzpicture}
\]
\caption{A theory with a Kramers-Wannier defect has a self-duality since the mesh of $\cC$ can be replaced by a set of topologically trivial loops of $\cN$. }
\label{fig:reversefig}
\end{center}
\end{figure}
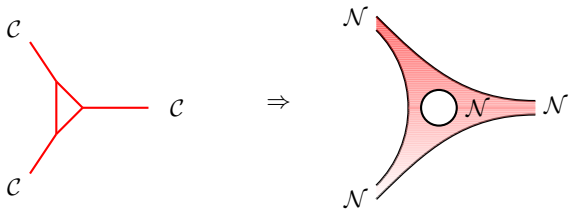

Conversely, assuming that we have a defect with fusion rules given in (\ref{eq:fusionrules}), there must be a corresponding self-duality. Indeed, we may begin by inserting a fine mesh of the condensate $\cC(M_3)$ (which is itself a fine mesh of surfaces), and then replacing it with pairs of $\cN(M_3)$ as shown schematically in Figure \ref{fig:reversefig}. But assuming that we started with a fine enough mesh, each loop of $\cN(M_3)$ is now contractible, and may be evaluated to a number. Thus with appropriate normalization we reobtain the original theory.

\section{Examples}
 We now give some examples of theories with Kramers-Wannier-type non-invertible defects, and hence with self-dualities.

\subsection{$SO(3)$ Yang-Mills theory at $\theta=\pi$}

As a first example, take the theory $\cT$ to be a pure $SU(2)$ Yang-Mills theory at $\theta=\pi$. This theory has a $\ZZ_2^{(1)}$ 1-form symmetry, as well as a time-reversal symmetry $\sT$.
These two symmetries are known to have an anomaly of the form (\ref{eq:mainanomaly}), with $A^{(1)}$ replaced by the first Stiefel-Whitney class $w_1^{TX_5}$ of the tangent bundle of $X_5$.\footnote{For $w_1^{TX_5}$ to be non-vanishing $X_5$ must be unorientable, but we will assume $X_4$ to be orientable and spin. The mixed anomaly on unorientable $X_4$ is more subtle and was discussed in \cite{Wan:2019oyr, Wang:2019obe, Wan:2018zql}. }  Our general construction tells us that upon gauging $\ZZ_2^{(1)}$, the codimension-1 defect implementing $\sT$ becomes non-invertible. Indeed, the resulting theory $\cT/\Z_2^{(1)}$ is $SO(3)$ Yang-Mills at $\theta = \pi$, which lacks the usual time-reversal symmetry since $\theta$ is $4 \pi$-periodic. Instead, it contains the non-invertible defect $\cN$ implementing a self-duality transformation under $TK S T$.

\begin{figure}[!tbp]
	\begin{center}
		\begin{tikzpicture}
		
		\draw[-stealth, thick] (0,0)--(3.2,0);
		\draw[-stealth, thick] (0,0)--(-3.2,0);
		\draw[ thick] (0,-0.1)--(0,0.1);
		\draw[ thick] (1.25,-0.1)--(1.25,0.1);
		\draw[ thick] (2.5,-0.1)--(2.5,0.1);
		\draw[ thick] (-1.25,-0.1)--(-1.25,0.1);
		\draw[ thick] (-2.5,-0.1)--(-2.5,0.1);
		\node[below] at (0,-0.1) {$0$};
		\node[below] at (-1.25,-0.1) {$-\pi$};
		\node[below] at (1.25,-0.1) {$\pi$};
		\node[below] at (2.5,-0.1) {$2\pi$};
		\node[below] at (-2.5,-0.1) {$-2\pi$};
		\node[right] at (3.2,0){$\,\,\theta$};
		\shade[line width=2pt, top color=red,opacity=0.4] 
		(-2.5,0) to [out=90, in=-90]  (-2.5,1.25)
		to [out=0,in=180] (-1.25,1.25)
		to [out = -90, in =90] (-1.25,0)
		to [out=190, in =0]  (-2.5,0);
		\shade[line width=2pt, top color=blue,opacity=0.4] 
		(-1.25,0) to [out=90, in=-90]  (-1.25,1.25)
		to [out=0,in=180] (1.25,1.25)
		to [out = -90, in =90] (1.25,0)
		to [out=190, in =0]  (-1.25,0);
		
		\shade[line width=2pt, top color=red,opacity=0.4] 
		(1.25,0) to [out=90, in=-90]  (1.25,1.25)
		to [out=0,in=180] (2.5,1.25)
		to [out = -90, in =90] (2.5,0)
		to [out=190, in =0]  (1.25,0);
		
		\node[above] at (0,0.3) {$\ZZ_2$ TQFT};
		\node[above] at (-1.875,0.7) {trivial};
		\node[below] at (-1.875,0.7) {TQFT};
		\node[above] at (1.875,0.7) {trivial};
		\node[below] at (1.875,0.7) {TQFT};
		\end{tikzpicture}\caption{Phase diagram of $SO(3)$ YM as a function of $\theta$. }
		\label{SO3phase}
	\end{center}
\end{figure}
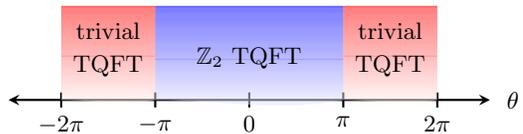

The non-invertible defect $\cN$ also suggests the structure of phases of $SO(3)$ Yang-Mills as a function of the theta angle. From the UV perspective, we find that
\begin{eqnarray}\label{Trans}
TK S T: (SO(3), \,\theta) \to (SO(3), \,2\pi-\theta)~.
\end{eqnarray}
Hence $\theta=\pm \pi\mod 4\pi$ is invariant under $TK S T$. At these points the defect implementing the transformation $TK S T$ between different theories at generic theta becomes one implementing self-duality of a single theory. This suggests that there should be a phase transition at these fixed values of theta. Indeed, just such a transition is expected on the basis of, e.g.\ soft supersymmetry breaking. See Figure \ref{SO3phase} for a schematic phase diagram. 

Let discuss this phase diagram further. In \cite{Aharony:2013hda} it was argued via soft supersymmetry breaking that for $ |\theta| <\pi$ the theory flows to a $\Z_2$ TQFT, while for $\pi<|\theta|<2\pi$, the theory flows to a trivially gapped phase. The phase transitions at $\theta=\pm \pi\mod 4\pi$ are transitions between these two low-energy phases. Our result is in agreement with \cite{Koide:2021zxj}, where a non-invertible defect was found in a lattice model which exhibits the same phase transition.

\subsection{$\cN=1$ $SO(3)$ super Yang-Mills theory}

Next we take $\cT$ to be $\cN=1$ $SU(2)$ super Yang-Mills. The symmetry of this theory is $\Z_4^{(0)}\times \Z_2^{(1)}$, where $\Z_4^{(0)}$ contains the fermion parity $\Z_2^F$ as a normal subgroup. There is a mixed anomaly between $\Z_4^{(0)}$ and  $\Z_2^{(1)}$ with the anomaly inflow action as in \eqref{eq:mainanomaly}, where $A^{(1)}$ is now the background field of $\Z_4^{(0)}$. Note that $\Z_2^F\subset \Z_4^{(0)}$ is anomaly free. The low-energy dynamics are well-known: the $\Z_4^{(0)}$ is spontaneously broken to $\Z_2^F$, and there are two gapped vacua related by $\Z_2^{(0)}:= \Z_4^{(0)}/\Z_2^F$. Each vacuum is trivially gapped. 

After gauging $\Z_2^{(1)}$, we obtain $\cN=1$ $SO(3)$ super YM.  Since $\Z_2^{(0)}$ is extended by a non-anomalous $\Z_2^F$, the fusion rules for the Kramers-Wannier duality defect discussed in the previous section are modified. Denote the 3d defect implementing $\Z_2^F$ by $F(M_3)$, and the $\Z_2^{(0)}$ defect before gauging  as $D(M_3)$. The fact that they together generate $\Z_4^{(0)}$ implies $D\times D=F$ and $F\times F=1$. Correspondingly, the fusion rule \eqref{eq:fusionrules} becomes 
\begin{equation}\label{NNext}
\begin{split}
&\cN(M_3)\times \cN(M_3)\\&\quad= {F(M_3) \over |H^0(M_3,\ZZ_2)|}  \sum_{\Sigma \in H_2(M_3, \ZZ_2)} (-1)^{Q(\Sigma)} L(\Sigma)~.
\end{split}
\end{equation}
The other fusion rules \eqref{eq.NLfusionrule} and \eqref{eq.LLfusionrule} are unmodified. Eq. \eqref{NNext} can be understood as an extension of the non-invertible symmetry by an invertible symmetry $\Z_2^F$.

There are still two vacua in the $SO(3)$ theory \cite{Aharony:2013hda}. One vacuum remains trivially gapped, while the other vacuum supports a nontrivial $\Z_2$ TQFT.  
Thus the two vacua are not exchanged by a conventional 0-form symmetry.
Instead, they are exchanged by acting with a non-invertible line $\cN$ implementing the self-duality $T S T$. The existence of the two vacua related by self-duality can be viewed as spontaneous breaking of the non-invertible symmetry.

\subsection{$\cN=4$ $SU(2)$ super Yang-Mills theory at $\tau=i$}

Finally, we note that $\cN=4$ $SO(3)_-$ super Yang-Mills theory at $\tau=i$ is invariant under $S$-duality, which effectively maps $\tau\to -1/\tau$ \cite{Aharony:2013hda}.  This $S$-duality is an invertible $\Z_2^{(0)}$ symmetry. There is also a $\Z_2^{(1)}$ 1-form symmetry, which has a mixed anomaly with $S$-duality
\begin{equation}
Z_{SO(3)_-}[-1/\tau, B^{(2)}]= e^{i \frac{\pi}{2}\int_{X_4} \cP(B^{(2)})} Z_{SO(3)_-}[\tau, B^{(2)}]~,
\end{equation}
following from $Z_{SU(2)}[\tau, B^{(2)}]= Z_{SO(3)_+}[-1/\tau, B^{(2)}]$.
 Hence the general results of the previous section apply here as well.

Concretely, we take $\cT$ to be $\cN=4$ $SO(3)_-$ super YM at $\tau=i$. Then $ TS\cT$ is $SU(2)$ super YM at $\tau=i$.  Our general results imply that $SU(2)$ super YM at $\tau=i$ contains a non-invertible defect $\cN$ implementing the Kramers-Wannier self-duality under {${S}$}. More generally, on the conformal manifold parameterized by $\tau= 2\pi i/g^2$, $SU(2)$ super YM theories come in pairs related by
\begin{eqnarray}
{S}: (SU(2), \,g) \to (SU(2), \,2\pi/g)~.
\end{eqnarray}
The theory at the fixed point $g=\sqrt{2\pi}$ is the 4d analogue of the topological transition studied in \cite{Ji:2019ugf}.

\section*{Acknowledgements}

We would like to thank Fabian Lehmann, Shu-Heng Shao, Sahand Seifnashri, Yuji Tachikawa, and Gabi Zafrir for discussions. 
J.K.\ would like to thank Kavli IPMU for their generous hospitality during the inception of this work. K.O.\ is supported by the Simons Collaboration on Global Categorical Symmetries.
 Y.Z.\ is partially supported by WPI Initiative, MEXT, Japan at IPMU, the University of Tokyo. 
\textbf{Note added:} While finalizing this work, we were informed that work on a similar topic will appear in \cite{othergroup}.

\bibliography{bib}

\appendix

\section{Explicit derivations of fusion rules}
\label{app:Lagrangian}
In this appendix we give an explicit derivation of the fusion rules of  $\mathcal{N}(M_3)$ and $L(\Sigma)$ defined in the main text. To do so, we make use of the continuum Lagrangian description of the defects. 
To begin, recall that the defect $\cN(M_3)$ is obtained by dressing $D(M_3, b^{(2)})$, the defect for the zero-form symmetry in the ungauged theory, with a $U(1)_2$ Chern-Simons theory. It can be written explicitly as 
\begin{equation}\label{NM3}
\cN(M_3)\propto \int \cD a \, D(M_3, b^{(2)}) \, e^{{i \over 2\pi} \int_{M_3} ada -i \int_{M_3} ab^{(2)}}~.
\end{equation} 
Here $a$ is a $U(1)$ dynamical field with normalization such that $\frac{1}{2\pi}d a$ is the local expression for the field strength valued in integers. 

We begin by determining the fusion rule $\cN\times \cN$.
As mentioned in the main text, the theory on this product is the double-semion theory, which is dual to $\ZZ_2$ gauge theory with a DW twist. To see this explicitly, we start from
\begin{eqnarray}\label{NN}
\begin{split}
&\cN(M_3)\times \cN(M_3) = \cA^{2,1}(M_3)\otimes \cA^{2,-1}(M_3)\\
&\quad\propto \int \cD a \cD a' e^{{i \over 2\pi} \int_{M_3} (ada-a'da') -i \int_{M_3} (a-a')b^{(2)}}
\end{split}
\end{eqnarray}
and change variables from $a'$ to $a-\widehat{a}$. The argument of the exponential then becomes 
\bea
\frac{2}{2\pi}ad\widehat{a}-\frac{1}{2\pi}\widehat{a}d\widehat{a}- \widehat{a}b^{(2)}~.
\eea
 Integrating out $a$ enforces that $\widehat{a}$ is locally trivial, i.e. $d\widehat{a}=0$, but there can be nontrivial $\Z_2$ holonomies $e^{i \oint \widehat{a}}=\pm 1$. Summing over $\widehat{a}$ then yields the result quoted in \eqref{eq:fusionrules}, up to overall normalization. The normalization is fixed in Appendix \ref{app:relcoho}.

Next we determine the fusion rule $\cN\times L$. Suppose that $\Sigma$ is a surface embedded in $M_3$. Then by definition we have
\begin{equation}\label{NL}
\begin{split}
&\cN(M_3)\times L(\Sigma)\\
&\quad=\int \cD a \, D(M_3, b^{(2)}) \, e^{{i \over 2\pi} \int_{M_3} ada -i \int_{M_3} ab^{(2)}+ i \pi \oint_{\Sigma} b^{(2)}}~ .
\end{split}
\end{equation}
One can rewrite the integral $\oint_{\Sigma}b^{(2)}= \int_{M_3} \delta^{\Sigma}b^{(2)}$, where $\delta^{\Sigma}$ is the Poincare dual of the surface $\Sigma$. One is free to redefine $a\to a+\pi \delta^{\Sigma}$ to absorb the last term since $a$ is path integrated, but upon doing so the Chern Simons term $\frac{i}{2\pi}\int ada$ becomes $\frac{i}{2\pi}\int ada + \frac{i\pi}{2}\int \delta^{\Sigma}d \delta^{\Sigma}$. This additional term measures the triple intersection number of $\Sigma$ mod 2, leading to the fusion rule \eqref{eq.NLfusionrule}.

%%%%%%%%%%%%%%%%%%%%%%%%%%%%%
%%%%%%%%%%%%%%%%%%%%%%%%%%%%%
\section{Normalization of $\cN$}
\label{app:relcoho}
%%%%%%%%%%%%%%%%%%%%%%%%%%%%%
%%%%%%%%%%%%%%%%%%%%%%%%%%%%%

In this appendix, we fix the overall normalization of the condensate $\cC(M_3)$ in (\ref{eq:condop}), and consequently of the fusion rule \eqref{eq:fusionrules}. To do so, note that the condensation operator $\cC(M_3)$ is obtained by considering a small tubular neighborhood around $M_3$, with the 1-form gauge theory inside and with Dirichlet boundary conditions imposed~\cite{Gaiotto:2019xmp}.

 In general, gauging a one-form symmetry $G$ on a closed $d$-manifold $X_d$ gives
\bea
Z_{\cT/G}[X_d] = \frac{|H^0(X_d,G)|}{|H^1(X_d,G)|}\sum_{B\in H^2(X_d,G)} \eps(B)\, Z_{\cT}[X_d,B],\qquad\,\,\,
\eea
where the normalization follows from dividing out the volume of the gauge group \cite{Gaiotto:2014kfa}, and $\eps(B)$ is a possible DW twist. We will ultimately be interested in the case with $d=4$, $G= \mathbb{Z}_2$, and $\eps(B) = e^{i {\pi \over 2} \int_{X_4} \cP(B)}$, but for the moment we remain more general.

We consider $X_d$ to be a tubular neighborhood of $M_{d-1}$, i.e. $X_d  = M_{d-1} \times I$ with $I=[0,1]$ an interval. This space has a boundary, and we impose Dirichlet boundary conditions on it. This means that we should work with relative cohomology equipped with the isomorphism
\bea
\label{eq:relcohomiso}
\Phi: H^*(X_d, \partial X_d, G)  \,\stackrel{\cong}{\rightarrow} \, H^{*-1}(M_{d-1}, G)~,
\eea
where $H^{-1}(M_{d-1}, G)$ is understood to be trivial. 

It will be useful to have an explicit form for this isomorphism, which we denote by $\Phi$. We set $*= k$ and restrict to $G=\ZZ_n$. In that case, each element can be locally written as a differential form $b^{(k)}$ satisfying
\bea
\label{eq:relcodef}
n \,d b^{(k)} = 0 \,,\hspace{0.25in} b^{(k)} |_{\p X_d}= 0~,
\eea
where we include $d b^{(k)} \in H^{k+1}(X_d,\partial X_d,\mathbb{Z})$ the torsion part coming from the global information of the $k$-gerbe $b^{(k)}$. For this reason we cannot divide the first equation of \eqref{eq:relcodef} by $n$.

We begin by decomposing $b^{(k)}$ as 
\bea
b^{(k)} = \omega^{(k)}(t) + dt \wedge \beta^{(k-1)}(t)~,
\eea
where $t$ is the coordinate on the interval $I$ and  $\omega^{(k)}$ is a $k$-form with no factors of $dt$. 
The second condition of (\ref{eq:relcodef}) requires that $\omega^{(k)}(0) = \omega^{(k)}(1) = 0$, and likewise for $\beta^{(k-1)}(t)$. The first condition of (\ref{eq:relcodef}) requires that 
\bea
n\, {\p \omega^{(k)} \over \p t} = n\, \mathrm{d}_M \beta^{(k-1)} ~, \hspace{0.25 in} n\, \mathrm{d}_M \omega^{(k)}  = 0  ,
\no\\
\eea
where $\mathrm{d}_M$ is the exterior derivative on $M_{d-1}$.

 The map in (\ref{eq:relcohomiso}) is then given by $\Phi: b^{(k)} \mapsto a^{(k-1)}$, where 
\bea
a^{(k-1)} := \int_0^1 dt \, \beta^{(k-1)}(t)~.
\eea
Note that this is indeed an element of $H^{k-1}(M_{d-1}, \mathbb{Z}_n)$. For example, closure on $M_{d-1}$ is seen via 
\bea
 n\, \mathrm{d}_M a^{(k-1)} &=& n\, \int_0^1 dt \,  \mathrm{d}_M\beta^{(k-1)}(t)= n\, \int_0^1 dt \,  {\p \omega^{(k)} \over \p t}
 \no\\
 & =&n\, (\omega^{(k)}(1)- \omega^{(k)}(0))= 0~.
 \label{eq:nda}
\eea
 It can further be proven that $\Phi$ is an isomorphism, though we will not do so here.

We now restrict to $d=4$, $k=2$ and ask for the image of the Pontryagin square under $\Phi$. To do so, we take $b^{(2)}$ and consider $b^{(2)} \wedge b^{(2)}$ under $\Phi$. Noting that, locally,
\bea
b^{(2)} \wedge b^{(2)}= 2 \,dt \wedge \omega^{(2)}  \wedge \beta^{(1)},
\eea
we get 
\bea
\Phi(b^{(2)} \wedge b^{(2)}) = 2 \int_0^1 dt \, \omega^{(2)}(t)  \wedge \beta^{(1)}(t)~.
\eea
To put this in a more appealing form, we define 
\bea
\tilde \beta^{(1)}(s) := \int_0^s dt\, \beta^{(1)}(t)
\eea
such that $\beta^{(1)}(s) = {\p \over \p s}\tilde \beta^{(1)}(s)$. Then we have 
\bea
\Phi(b^{(2)} \wedge b^{(2)}) &=&  2 \int_0^1 dt \, \omega^{(2)}(t)  \wedge {\p \over \p t} \tilde \beta^{(1)}(t)
\no\\
&=& -2 \int_0^1 dt \,  {\p \over \p t} \omega^{(2)}(t)  \wedge \tilde \beta^{(1)}(t)
\no\\
&=& 2 \int_0^1 dt \, \mathrm{d}_M \beta^{(1)}(t)  \wedge \tilde \beta^{(1)}(t)~.
\eea
There is no boundary term from the integration by parts since $\omega^{(2)}$ vanishes on the boundary. Finally, we note that 
\bea
\Phi(b^{(2)} \wedge b^{(2)}) &=& 2 \int_0^1 dt \, \mathrm{d}_M \beta^{(1)}(t)  \wedge \int_0^t ds\, \beta^{(1)}(s)
\no\\
&=& \int_0^1 dt ds \, \beta^{(1)}(s) \wedge \mathrm{d}_M \beta^{(1)}(t) 
\no\\
&=& a^{(1)} \wedge \mathrm{d}_M a^{(1)} ~.
\eea
Note that this is not necessarily 0, as we cannot divide \eqref{eq:nda} by $n=2$.
In cocycle notation, this reads
\bea
\Phi(\cP(b^{(2)})) = 2\, a^{(1)} \cup a^{(1)} \cup a^{(1)} \,\,\, \mathrm{mod}\,\, 4~.
\eea

To summarize, we see that 1-form gauge theory with Dirichlet boundary conditions on a tubular neighborhood of $M_3$ and with Dijkgraaf-Witten twist $e^{i {\pi \over 2} \int_{X_4} \cP(B)}$ gives, upon application of $\Phi$,
\begin{equation}
\label{eq:finalgauging}
Z_{\cT/G}[M_3] = \frac{1}{|H^0 (M_3,G)|}\sum_{a\in H^1(M_3,G)} (-1)^{Q(\mathrm{PD}(a))} Z_{\cT}[M_3,a], 
\end{equation}
with $\mathrm{PD}(a)$ the Poincar{\'e} dual to $a$ and $Q$ the triple intersection number. The condensation operator $\cC(M_3)$, i.e. the operator implementing the gauging, then takes the form shown in (\ref{eq:condop}). 

The fact that (\ref{eq:finalgauging}) looks like a 0-form gauge theory (as opposed to the naive expectation of a 1-form gauge theory) on $M_3$ is because the 1-form symmetry defects are codimension-2 in $X_4$, but codimension-1 in $M_3$.

%%%%%%%%%%%%%%%%%%%%%%%%%%%%%
%%%%%%%%%%%%%%%%%%%%%%%%%%%%%
\section{Non-invertible defects in (3+1)d from $\Z_N^{(0)}\times \Z_M^{(1)}$ mixed anomaly}
\label{app:ZNZM}
%%%%%%%%%%%%%%%%%%%%%%%%%%%%%
%%%%%%%%%%%%%%%%%%%%%%%%%%%%%

In this appendix, we consider the non-invertible defects arising from gauging the $\Z_M^{(1)}$ symmetry of a theory $\cT$ with a mixed anomaly between $\Z_N^{(0)}$ and $\Z_M^{(1)}$. The special case $N=M=2$ reduces to the situation discussed in the main text.

Suppose that a (3+1)d theory $\cT$ has global symmetry $\Z_N^{(0)}\times \Z_M^{(1)}$, with respective background fields $A^{(1)}$ and $B^{(2)}$. Their mixed anomaly is captured by a 5d invertible TQFT
\begin{eqnarray}
\frac{2\pi}{\gcd(N,M)} \int_{X_5} A^{(1)} \cup {\cP (B^{(2)} )\over 2}~.
\end{eqnarray}
The anomaly implies that the $\Z_N^{(0)}$ invertible symmetry defect has a $\Z_M^{(1)}$ one-form symmetry anomaly, and the following combination is gauge invariant,
\begin{eqnarray}\label{DNM}
D(M_3, B^{(2)}) \,e^{\frac{2\pi i}{\gcd(N,M)} \int_{M_4} \cP(B^{(2)})/2},
\end{eqnarray}
with $\partial M_4 = M_3$. The $\Z_M^{(1)}$ symmetry has an anomaly labeled by $p= M/\gcd(N,M)\mod M$. For convenience, we will denote $K=\gcd(p,M)$.

We are interested in the theory $\cT/\Z_M^{(1)}$ obtained by gauging the $\Z_M^{(1)}$ symmetry of $\cT$. 
Gauging $\Z_M^{(1)}$ can be divided into a two step process, following \cite{Hsin:2018vcg}:
\begin{enumerate}
	\item Gauge $\Z_{K}^{(1)}$.  Since the normal subgroup $\Z_{K}^{(1)}$ is anomaly free, we can gauge $\Z_{K}^{(1)}$ and $D(M_3)$ is still a well-defined genuine 3d defect. Eq. \eqref{DNM} becomes
	\begin{eqnarray}\label{DMK}
	D(M_3, M\tilde{b}/K+ \tilde{B})\, e^{2\pi i \frac{p/K}{M/K}\int_{M_4}\cP(\tilde{B})/2}
	\end{eqnarray}
	where $\tilde{b}$ is the dynamical gauge field for $\Z_K^{(1)}$, and $\tilde{B}$ is the background gauge field $\Z_{M/K}^{(1)}$.\footnote{In $M\tilde{b}/K+ \tilde{B}$, the background field $\tilde{B}$ should be interpreted as its lift to $\Z_M^{(1)}$. To simplify notation, we use $\tilde{B}$ for both the $\Z_{M/K}^{(1)}$ background field and its lift.} The defect \eqref{DMK} is still invertible. 
	\item Gauge the remaining $\Z_{M/K}^{(1)}$. We need to couple a TQFT with anomaly opposite to that of \eqref{DMK}, i.e. $e^{-2\pi i \frac{p/K}{M/K}\int_{M_4}\cP(\tilde{B})/2}$, to make $D(M_3, M\tilde{b}/K+ \tilde{B})$ well-defined after gauging $\Z_{M/K}^{(1)}$. Since  $\gcd(p/K, M/K)=1$, any TQFT can be factorized into the decoupled product of a minimal TQFT $\cA^{M/K, -p/K}$ and another TQFT which is neutral under $\Z_{M/K}^{(1)}$.
\end{enumerate}
Combining the two steps, we find that after gauging $\Z_M^{(1)}$  the invertible defect \eqref{DNM} becomes a non-invertible defect 
\begin{eqnarray}
\cN(M_3)= D(M_3, b^{(2)})\, \cA^{M/K, -p/K}(M_3, b^{(2)})~,
\end{eqnarray}
where $\cA^{M/K, -p/K}(M_3, b^{(2)})$ only depends on $b^{(2)}$ mod $M/K$.

We next discuss the fusion rules. Consider the case when $\Z_N^{(0)}$ is anti-linear. When $N=2$, this symmetry can be taken to be time reversal, whereas more generally for $N=2n$ it can be an extension of time reversal by a $\Z_n$ symmetry, with $\sT^2=U$ and $U$ the generator of $\Z_n$. We will focus on this case, for which we find $\cN\times \cN= D\cA^{M/K, -p/K} \times D\cA^{M/K, -p/K}=U \cA^{M/K, p/K}\cA^{M/K, -p/K}$, where we have used $D^2=U$. We may further use the duality \cite{Hsin:2018vcg}
\begin{eqnarray}
\cA^{M/K, p/K}\cA^{M/K, -p/K} \leftrightarrow (\cZ_{M/K})_{-Mp/K^2}~,
\end{eqnarray}
which holds because $\gcd(M/K, p/K)=1$.  The notation $(\cZ_{M/K})_{-Mp/K^2}$ on the right-hand side denotes $\Z_{M/K}$ gauge theory with DW twist $-\frac{Mp/K^2}{4\pi}\int_{M_3} ada = - \pi p/K \int_{M_3} \tilde{a} \beta \tilde{a}$, where we have defined $\tilde{a}\in Z^{1}(M_3, \Z_{M/K})$ related to the original $U(1)$ gauge field $a$ via $a= \frac{2\pi}{M/K}\tilde{a}$.  The fusion rule then becomes 
\begin{equation}\label{fusionrulegeneral}
\begin{split}
&\cN(M_3)\times \cN(M_3)\\&= \frac{U(M_3)}{|H^0(M_3, \Z_{M/K})|}  \sum_{\Sigma\in H_2(M_3, \Z_{M/K})} (-1)^{\frac{p}{K} Q'(\Sigma)}L(\Sigma)
\end{split}
\end{equation}
where $L(\Sigma)= e^{\frac{2\pi i}{M/K}\int_{\Sigma} b^{(2)}}$. The factor $(-1)^{Q'(\Sigma)}$ is defined to be $(-1)^{\int_{M_3} \tilde{a}\beta \tilde{a}}$, with $\tilde{a}$ the Poincare dual of $\Sigma$ and $\beta$ the Bockstein operation. 
When $N=M=2$, \eqref{fusionrulegeneral} reduces to \eqref{eq:fusionrules}.

When $\Z_N^{(0)}$ is linear, we instead consider the fusion between two different operators: $\bar\cN\times \cN$, where $\bar\cN$ the orientation reversal of $\cN$. Explicitly, $\bar\cN= \bar D \cA^{M/K, p/K}$. Then the fusion rule becomes
\begin{equation}
\begin{split}
&\bar \cN(M_3)\times \cN(M_3)\\&= \frac{1}{|H^0(M_3, \Z_{M/K})|}  \sum_{\Sigma\in H_2(M_3, \Z_{M/K})} (-1)^{\frac{p}{K} Q'(\Sigma)}L(\Sigma)~.
\end{split}
\end{equation}

Using the above general results, it is possible to identify more (3+1)d gauge theories with non-invertible symmetries. For example, we may consider $SU(M)$ Yang-Mills with $\theta=\pi$, for which $N=2$ and $\Z_2^{(0)}$ is anti-linear.  Indeed, it is known that this theory is time reversal symmetric, and has a mixed anomaly with the $\Z_M^{(1)}$ one-form symmetry when $M$ is even \cite{Gaiotto:2017yup}. After gauging $\Z_M^{(1)}$, we obtain $PSU(M)$ Yang-Mills theory with $\theta=\pi$. This theory is not time reversal symmetric, but has a non-invertible topological defect.

%%%%%%%%%%%%%%%%%%%%%%%%%%%%%
%%%%%%%%%%%%%%%%%%%%%%%%%%%%%
\section{Non-invertible defects in (2+1)d}
\label{app:3d}
%%%%%%%%%%%%%%%%%%%%%%%%%%%%%
%%%%%%%%%%%%%%%%%%%%%%%%%%%%%

In this appendix we consider non-invertible defects of (2+1)d theories. The naive analog of the construction in the main text would be to consider a theory with anomaly 
\bea
\label{eq:possibleanom1}
\pi \int_{X_4} B^{(2)} \cup A^{(1)} \cup A^{(1)}
\eea
with $B^{(2)}$ and $A^{(1)}$ the background gauge fields for $\ZZ_2$ 1- and 0-form symmetries,  in analogy to (\ref{eq:mainanomaly}). We would then hope to gauge the 0-form symmetry to make the 1-form symmetry defect non-invertible. However, this approach will not work---indeed, gauging the 0-form symmetry does not give rise to non-invertible defects, but rather extends $\Z_2^{(1)}$ to $\Z_4^{(1)}$, due to the fact that (\ref{eq:possibleanom1}) is equivalent to 
\bea
\pi \int_{X_4} B^{(2)} \cup {\mathrm{Sq}^1}  A^{(1)}= \pi \int_{X_4} A^{(1)} \cup {\mathrm{Sq}^1}  B^{(2)}~.\,\,\,\,
\eea

To obtain Kramers-Wannier-like defects in (2+1)d, we instead begin with a theory $\cT$ with a $\ZZ_2$ 1-form symmetry and \textit{two} $\ZZ_2$ 0-form symmetries, with anomaly
\bea
\label{eq:good3danom}
\vphantom{.}\pi \int_{X_4} B^{(2)} \cup  A_1^{(1)} \cup A_2^{(1)} 
\eea
involving all three symmetries. For simplicity, we assume all of them to be linear.

\subsection{Non-invertible line and surface defects from gauging}
\label{app.noninvertiblelinesurface}

The presence of the anomaly \eqref{eq:good3danom} implies that the codimension-1 defect $D_1(M_2)$ of one of the $\Z_2^{(0)}$ symmetries  (call it $\ZZ_{2,1}^{(0)}$ with background field $A_1^{(1)}$) carries an anomaly under $\Z_2^{(1)}\times \Z_{2,2}^{(0)}$, with $\Z_{2,2}^{(0)}$ the remaining 0-form symmetry.  Only the following combination is invariant under background field transformations of $B^{(2)}$ and $A_2^{(1)}$, 
\begin{eqnarray}
D_1(M_2, B^{(2)}, A_2^{(1)}) \,e^{i \pi \int_{M_3} B^{(2)}\cup A_2^{(1)} }~.
\end{eqnarray}
As we will find below, gauging $\Z_2^{(1)}\times \Z_{2,2}^{(0)}$ by promoting $B^{(2)}$ and $A_2^{(1)}$ to dynamical fields $b^{(2)}$ and $a_2^{(1)}$ respectively will render the $\Z_{2,1}^{(0)}$ defect non-invertible.

 Indeed, gauging $\Z_2^{(1)}\times \Z_{2,2}^{(0)}$ makes $D_1(M_2, B^{(2)}, A_2^{(1)})$ ill-defined, and to restore gauge-invariance we must dress it with an appropriate worldsheet theory,
\bea
\cN_1(M_2) &\propto& \int \cD \phi_2^{(0)} \cD \eta^{(1)}\, D_1(M_2,, b^{(2)}, a_2^{(1)})\times
\no\\
&\vphantom{.}&  e^{i  \pi \int_{M_2} \phi_2^{(0)} b^{(2)} + \eta^{(1)}  a_2^{(1)} - \phi_2^{(0)} d  \eta^{(1)}}~,
\eea
with $\phi_2^{(0)}\in C^0(M_2, \Z_2)$ and $\eta^{(1)}\in  C^1(M_2, \Z_2)$.  These fields transform under $\Z_2^{(1)}\times \Z_{2,2}^{(0)}$ via shifts $\phi_2^{(0)}\to \phi_2^{(0)}+ \lambda_2^{(0)}$ and $\eta^{(1)}\to \eta^{(1)}+ \lambda^{(1)}$.  Following similar steps as in Appendix \ref{app:Lagrangian}, it is straightforward to compute the self-fusion of $\cN_1(M_2)$, with the result being
\begin{equation}\label{mixedfusionrule}
\cN_1(M_2) \times \cN_1(M_2) = {1+ W(M_2)\over |H^0(M_2, \ZZ_2)|} \sum_{M_1 \in H_1(M_2, \ZZ_2)} L_2(M_1)
\end{equation}
with $W(M_2) := e^{i \pi \oint_{M_2} b^{(2)}}$ and $L_2(M_1) :=e^{i \pi \oint_{M_1} a_2^{(1)}}$. 
The normalization has been fixed by considerations similar to those in Appendix \ref{app:relcoho}. Since there are multiple lines appearing on the right-hand side, $\cN(M_1)$ is non-invertible. 

The fusion rules between $\cN_1(M_2)$ and the other defects are also easily computed, giving
\bea\label{NWL}
\cN_1(M_2) \times W(M_2) &=& \cN_1(M_2)~,
\no\\
\cN_1(M_2) \times L_2(M_1) &=& \cN_1(M_2)~,
\eea
for any $M_1 \subset M_2$. Thus as desired, we find a Kramers-Wannier-like defect in (2+1)d.

In the next subsection, we will argue that this Kramers-Wannier-like defect implements a self-duality. However, before doing so, let us note that rather than gauging $\Z_2^{(1)}\times \Z_{2,2}^{(0)}$ of $\cT$ to find a non-invertible surface defect of $\cT/(\Z_2^{(1)}\times \Z_{2,2}^{(0)})$, we could have instead gauged $\Z_{2,1}^{(0)}\times \Z_{2,2}^{(0)}$ of $\cT$ to find a non-invertible line of $\cT/(\Z_{2,1}^{(0)}\times \Z_{2,2}^{(0)})$. Indeed, if before gauging we denote the defect line for the $\Z_2^{(1)}$ symmetry as $D(M_1)$, we see that this line supports an anomaly with the following combination being gauge invariant,
\begin{equation}
D(M_1,A_1^{(1)},A_2^{(1)}) \,e^{i \pi \int_{M_2} A_1^{(1)} \cup A_2^{(1)} } ~,
\end{equation}
where $M_1=\partial M_2$. After gauging $\Z_{2,1}^{(0)}\times \Z_{2,2}^{(0)}$, the $\Z_2^{(1)}$ defect line becomes the following non-invertible line,
\begin{eqnarray}
\begin{split}
\cN(M_1) \propto& \int \cD\phi_1^{(0)} \cD\phi_2^{(0)}  \,\,D(M_1,a_1^{(1)},a_2^{(1)}) \times 
\\
&\hspace{0.2 in}e^{i  \pi \int_{M_1} \phi_1^{(0)} a_2^{(1)} -\phi_2^{(0)} a_1^{(1)} - \phi_1^{(0)} d \phi_2^{(0)}  },
\end{split}
\end{eqnarray}
where $\phi_i^{(0)}\in C^0(M_1, \Z_2)$ and they transform under $\Z_{2,1}^{(0)}\times \Z_{2,2}^{(0)}$ as $\phi_i^{(0)}\to \phi_i^{(0)}+\lambda_i^{(0)}$. By following the steps in Appendix \ref{app:relcoho}, the fusion rules are found to be
\begin{equation}\label{eq:3dfusion}
\begin{split}
\cN(M_1)\times \cN(M_1)  &=  \prod_{i=1}^2 (1 + L_i(M_1) )~,\\
\cN(M_1) \times L_i(M_1) &= \cN(M_1) ~,\hspace{0.2 in} i=1,2~,
\end{split}
\end{equation}
where $L_i(M_1) :=e^{i \pi \oint_{M_1} a_i^{(1)}}$. Hence we now get a line with Kramers-Wannier-like fusion rules. Note that this is now a codimension-2 defect, and hence it no longer acts as a domain wall implementing self-duality.

Instead, the interpretation is as follows. In the gauged $\cT/(\Z_{2,1}^{(0)}\times \Z_{2,2}^{(0)})$ theory,
the codimension 1-defect $e^{i \pi \int_{\Sigma_2} a_1^{(1)}  a_2^{(1)}}$ on a compact $\Sigma_2$ is trivial because the anomaly of $\cT$ implies
\begin{equation}
	Z_{\cT}[M_3] = Z_{\cT}[M_3]e^{i \pi \int_{\Sigma_2} a_1^{(1)} a_2^{(1)}}.
\end{equation}
However, when $\Sigma_2$ is bounded by $M_1$ this does not necessarily hold, and the boundary $M_1=\partial \Sigma_2$ should support degrees of freedom for the operator to be gauge-invariant, which should be identified with the $\cN(M_1)$ operator. Just as $\cN(M_3)$ in the main text implements the condensation of $L(M_2)$ (with appropriate twists) in $\cT/\Z_2^{(1)}$ (see Fig.~\ref{fig:KWfusion}), $\cN(M_1)$ here implements the condensation of $e^{i\pi \int a_i^{(1)}}, i=1,2$ on $\Sigma_2$ which is bounded by $M_1$.

\subsection{Non-invertible surface defect $\cN_1$ implementing self-duality}

We now return to the codimension-1 defect $\cN_1(M_2)$ and show that it implements a self-duality similar to that in the (3+1)d theories discussed in the main text.

The partition functions of the theory $\cT$ and its gauging $\cT':= \cT/(\Z_2^{(1)}\times \Z_{2,2}^{(0)})$ are related via 
\begin{eqnarray}
\begin{split}
Z_{\cT'}[\widehat{B}^{(1)}, \widehat{A}_2^{(2)}]= \int &\cD b^{(2)} \cD a_2^{(1)} Z_{\cT}[a_2^{(1)}, b^{(2)}] \times \\&\hspace{0.1in}e^{i \pi \int_{X_3}b^{(2)}\widehat{B}^{(1)} + a_2^{(1)}\widehat{A}_2^{(2)}}~.
\end{split}
\end{eqnarray}
We now show that $\cT'$ is invariant under the transformation $TST$.  The operation $T$ is defined to be addition of a counterterm $e^{i \pi \int \widehat{B}^{(1)}\widehat{A}_2^{(2)}}$, while $S$ is defined as gauging $\widehat{\Z}_{2,2}^{(1)}\times \widehat{\Z}_{2}^{(0)}$. Under $ST$, the partition function of $\cT'$ becomes 
\begin{equation}\label{beomes}
\begin{split}
&\int \cD \widehat{b}^{(1)} \cD \widehat{a}_2^{(2)} Z_{\cT'}[\widehat{b}^{(1)}, \widehat{a}_2^{(2)}] e^{i \pi \int_{X_3} \widehat{b}^{(1)} \widehat{a}_2^{(2)} +\widehat{b}^{(1)} \widetilde{B}^{(2)} +  \widehat{a}_2^{(2)} \widetilde{A}_2^{(1)} }\\
&\,\,\,\,=\int \cD b^{(2)} \cD a_2^{(1)} Z_{\cT}[a_2^{(1)}, b^{(2)}] e^{i \pi \int_{X_3} (b^{(2)}+ \widetilde{B}^{(2)}) (a_2^{(1)} + \widetilde{A}_2^{(1)})}~.
\end{split}
\end{equation}

We further use the fact that $\cT$ is invariant under $\Z_{2,1}^{(0)}$ and has the mixed anomaly \eqref{eq:good3danom}, which implies that
\begin{eqnarray}
Z_{\cT}[a_2^{(1)}, b^{(2)}]= Z_{\cT}[a_2^{(1)}, b^{(2)}]\, e^{i \pi \int_{X_3} a_2^{(1)} b^{(2)}}~,
\end{eqnarray}
i.e. $Z_{\cT}[a_2^{(1)}, b^{(2)}]$ is zero unless $e^{i \pi \int_{X_3} a_2^{(1)} b^{(2)}}$ is trivial. 
Then \eqref{beomes} reduces to 
\begin{eqnarray}
Z_{\cT'}[\widetilde{A}_2^{(1)}, \widetilde{B}^{(2)}] \,e^{i\pi \int_{X_3} \widetilde{A}_2^{(1)} \widetilde{B}^{(2)}}~,
\end{eqnarray}
where the counterterm can be removed by a further $T$ transformation. Hence $\cT'$ is indeed self-dual under $TST$, analogous to the (3+1)d case in the main text.

One can further argue that the self-duality $TST$ is implemented via the non-invertible line $\cN_1(M_2)$. The discussion is almost identical to that around Figure \ref{fig:KWfusion} in the main text. In the current case the $S$ transformation implements a gauging of $\widehat{\Z}_{2,2}^{(1)}\times \widehat{\Z}_{2}^{(0)}$, so between the two $\cN_1$ defects is a fine mesh of $W$ and $L_i$ defects.

However, unlike in the (3+1)d case, we notice that now the first $T$ operation does not contribute a nontrivial phase factor (i.e. an analog of $(-1)^{Q(\Sigma)}$) to the fusion rules. This can be confirmed by using the map $\Phi$ defined in \eqref{eq:relcohomiso} of Appendix \ref{app:relcoho}. Indeed, let us denote the images of $\hat b^{(1)}$ and $\hat a_2^{(2)}$ under the map $\Phi$ by 
\bea
 \beta^{(0)}:= \Phi(\hat b^{(1)}) ~, \hspace{0.2 in}\alpha^{(1)}:= \Phi(\hat a_2^{(2)}) ~.
\eea
Then one finds that 
\bea
\Phi(\hat b^{(1)} \cup \hat a_2^{(2)}) \,=\, 2 \, \beta^{(0)} \,  \alpha^{(1)} \cup  \alpha^{(1)}
\eea
which vanishes mod 2.

\subsection{Examples}

We now provide some examples of non-invertible surfaces and lines in (2+1)-dimensions.
\subsubsection{Non-invertible lines in Chern-Simons-matter theories}
\label{CMSlines}

We start by taking $\cT$ to be  the Chern-Simons-matter theory
\begin{eqnarray}\label{SOCSM}
SO(N)_K ~\text{with}~ N_f~\text{adjoint scalars}~.
\end{eqnarray}
We will restrict to the case of $N,K,N_f=0\mod 4$. Such theories have 1-form center symmetry $\Z_2^{(1)}$, 0-form charge conjugation symmetry $\Z_2^\sC$ and 0-form magnetic symmetry $\Z_2^\sM$. Since adding a mass term for all the flavors does not break any of the above symmetries, the `t Hooft anomalies can be found from the pure $SO(N)_K$ Chern-Simons theory obtained by large positive (degenerate) mass deformation. The anomaly of the  1-form symmetry is $\frac{\pi NK}{4}\int \cP(B^{(2)})/2$, which vanishes on arbitrary manifolds, but there is a nontrivial mixed anomaly
\begin{eqnarray}
\pi \int_{X_4} B^{(2)}\cup A^\sC\cup A^\sM
\end{eqnarray}
which is of the same form as \eqref{eq:good3danom} \cite{Cordova:2017vab}. One could also use the large negative mass deformation to obtain $SO(N-N_f)_K$, with the anomalies remaining the same.  

We can now apply the general discussion in the previous subsections here. After gauging $\Z_2^\sC\times \Z_2^\sM$, the theory becomes
\begin{eqnarray}\label{Pin+theory}
Pin^+(N)_K~\text{with}~ N_f~\text{adjoint scalars}
\end{eqnarray}
where adjoint means that the scalar transforms in the adjoint representation of the $\so(N)$ Lie algebra. The general discussion of the previous subsections shows that \eqref{Pin+theory} has a non-invertible topological line satisfying the fusion rules (\ref{eq:3dfusion}).

Note that the non-invertible topological lines appear in both the pure Chern-Simons theory $Pin^+(N)_K$, as well as the Chern-Simons-matter theory with or without mass. In the former case, the non-invertible topological lines are simply anyons which are ubiquitous for generic 3d TQFTs. However, the presence of non-invertible topological lines in the latter is nontrivial. Indeed, it is typically expected that upon coupling to matter and turning on finite Yang-Mills couplings, the topological lines of Chern Simons theories become non-topological, unless they are the generators of conventional 1-form symmetries (which are invertible). We now see that this is not entirely true: there exist non-invertible lines which remain topological even after deforming away from the TQFT limit!

The same conclusion holds if we replace the adjoint scalars by fermions in an appropriate representation, with Chern-Simons levels suitably adjusted.  We provide two examples:
\begin{enumerate}
	\item Take $\cT$ to be the Chern-Simons-matter theory $SO(N)_{K+\frac{(N-2)N_f}{2}}$ with $N_f$ Majorana fermions in the adjoint representation, with $N, K, N_f=0\mod 4$.  Gauging $\Z_2^\sC\times \Z_2^\sM$ gives
	$Pin^+(N)_{K+\frac{(N-2)N_f}{2}}$ with $N_f$ adjoint Majorana fermions, 
	which also has a non-invertible topological line satisfying the fusion rules  (\ref{eq:3dfusion}).
	\item Take $\cT$ to be Chern Simons matter theory $SO(N)_{K+\frac{(N+2)N_f}{2}}$ with $N_f$ Majorana fermions in the two-index symmetric representation, again with $N, K, N_f=0\mod 4$.  Gauging $\Z_2^\sC\times \Z_2^\sM$ gives
	$
	Pin^+(N)_{K+\frac{(N+2)N_f}{2}}$ with $N_f$ $S$  Majorana fermions
	where $S$ represents the two index symmetric representation of $\so(N)$. This theory again has non-invertible topological lines. 
\end{enumerate}

It would be nice to see if the non-invertible lines can be matched across Chern-Simons-matter dualities. However, as far as we know, Chern-Simons-matter dualities for $N_f\in 4\Z$ flavors of adjoint or symmetric tensor matter have not been studied in the literature (but see \cite{Gomis:2017ixy,Cordova:2017vab,Choi:2018tuh} for dualities involving $N_f=1$ matter, and \cite{Choi:2019eyl} for dualities involving $N_f=2$ matter).  It would be interesting to study this in the future.

\subsubsection{Non-invertible surfaces in Chern-Simons-matter theories}

We now begin with the Chern-Simons-matter theory \eqref{SOCSM} and gauge $\Z_2^{(1)}\times \Z_2^\sM$ instead of $\Z_2^\sC\times \Z_2^\sM$. Note that gauging the $\Z_2^\sM$ of $SO(N)$ gauge theory yields $Spin(N)$ gauge theory. The $Spin(N)$ gauge theory has $\Z_2^{(1)}\times \widehat{\Z}_2^{(1)}$ one-form symmetry, with $\text{diag}(\widehat{\Z}_2^{(1)}, \Z_2^{(1)})$ being the quantum symmetry arising from gauging $\Z_2^\sM$.\footnote{Denote the generators of $\Z_2^{(1)}$ and $\widehat{\Z}_2^{(1)}$  by $x$ and $y$ respectively. The generator of $\text{diag}(\widehat{\Z}_2^{(1)}, \Z_2^{(1)})$ is $xy$. Gauging the $\text{diag}(\widehat{\Z}_2^{(1)}, \Z_2^{(1)})$ symmetry of $Spin(N)$ gauge theory gives back the $SO(N)$ gauge theory.} Further gauging $\Z_2^{(1)}$ yields $Ss(N)$  gauge theory \cite{Aharony:2013hda}. Applying the result from the previous subsections, we find  
\begin{eqnarray}\label{Sstheory}
Ss(N)_K~\text{with}~ N_f~\text{adjoint scalars}
\end{eqnarray}
has a non-invertible surface $\cN_1(M_2)$ satisfying the fusion rules \eqref{mixedfusionrule} and \eqref{NWL}. Moreover, \eqref{Sstheory} enjoys a self duality under $TST$. 

One can also gauge $\widehat{\Z}_2^{(1)}$ to find $Sc(N)$ gauge theory.  Applying the result from the previous subsections, we find that  
\begin{eqnarray}\label{Sctheory}
Sc(N)_K~\text{with}~ N_f~\text{adjoint scalars}
\end{eqnarray}
has a non-invertible surface $\cN_1(M_2)$ satisfying the fusion rules \eqref{mixedfusionrule} and \eqref{NWL}. Moreover, \eqref{Sctheory} enjoys a self duality under $TST$. 

One can similarly discuss theories with fermions in either the adjoint or two-index symmetric tensor representation. The discussion is similar to that in Appendix \ref{CMSlines}, and hence we will not repeat it here.

\section{ (1+1)d Kramers-Wannier duality}
\label{app:2dcase}
In this appendix we show that the usual (1+1)d Kramers-Wannier line and self-duality can be understood in a way analogous to that in the main text. For related discussions, see \cite{Karch:2019lnn,Lin:2019hks}.

\subsection{Non-invertible defect}

The starting point is the theory $\cT$ of a single Majorana fermion. This theory has a pair of 0-form symmetries, denoted $(-1)^{\mathsf{F}}$ and $(-1)^{\mathsf{F_L}}$, which are respectively total and left-moving fermion parities. 
These symmetries have a mixed `t Hooft anomaly, which implies that the partition function on $X_2$ with the spin structure $\cS$ can absorb the Arf-invariant \cite{Witten:2015aba},
\begin{eqnarray}\label{2danom}
Z[X_2, \cS]\,\, =\,\, Z[X_2, \cS]\, (-1)^{\mathrm{Arf}(X_2,\cS)}~.
\end{eqnarray}
In other words, one finds $Z[X_2, \cS]=0$ whenever $(-1)^{\mathrm{Arf}(X_2,\cS)}=-1$. When $X_2=T^2$, this is manifested by the zero mode that is present for $RR$ spin structure. The mixed anomaly also implies that the topological defect line $D(M_1, \cS)$ corresponding to $(-1)^{\mathsf{F_L}}$ can be made invariant under background transformations by coupling it to a bulk term
\bea\label{DArf}
D(M_1, \cS) \,e^{i \pi\mathrm{Arf}(M_2,\cS)}
\eea
with $\p M_2 = M_1$.

To gauge $(-1)^{\mathsf{F}}$ we replace $\cS$ by $\cS+ \lambda$, with $\lambda$ a dynamical $\ZZ_2$ gauge field which should be integrated over. Since everything now depends only on the combination $\cS+\lambda$ (rather than $\cS$ and $\lambda$ independently), to simplify notation we will pick a spin structure $\cS$ such that $(-1)^{\mathrm{Arf}(\cS)}=1$ and suppress $\cS$ throughout. The $(-1)^{\mathsf{F_L}}$ defect after gauging then becomes $D(M_1,\lambda)$. From \eqref{DArf} this is no longer gauge-invariant, and hence it must be dressed with an appropriate 1d theory. Indeed, we may define the gauge invariant line

\bea
\cN(M_1):= D(M_1, \lambda) \,\cM(M_1,\lambda)~
\eea
with $\cM(M_1,\lambda)$ a Majorana zero mode \cite{kitaev2001unpaired,Kapustin:2014dxa}. When we consider the product of two $\cN(M_1)$, there are a pair of Majorana zero modes on the resulting line, and this pair admits a consistent quantization with Hilbert space of dimension 2. Recalling that the spinor representation $\rho_\psi$ fuses via $\rho_\psi \times \rho_\psi =\rho_0 \oplus \rho_{\mathrm{vec}}$, we conclude that we must have the fusion rules quoted in (\ref{eq.originalKW}).

\subsection{Self-duality}

We now show that the self-duality in this theory is of the same form $TST$ as in the main text. Here $S$ represents gauging of $\hat \ZZ_2^{(0)}$, and $T$ represents addition of an Arf twist. To show this, first note that 
\bea
Z_{\cT/\ZZ_2^{(0)}}[\Omega] = {1\over \sqrt{b_1}}\sum_{\lambda \in H^1(X_2, \ZZ_2)} Z_{\cT}[\lambda] (-1)^{\int_{X_2} \lambda \cup \Omega}~,
\eea
where $b_1 :=|H^1(X_2, \ZZ_2)|$. 
Then upon doing $T$ and $S$ transformations, we have 
\bea
&\vphantom{.}& {1\over \sqrt{b_1}}\sum_{\omega\in H^1(X_2, \ZZ_2) }Z_{\cT/\ZZ_2^{(0)}}[\omega] (-1)^{\mathrm{Arf}( \omega)+ \int_{X_2} \omega \cup \Lambda} 
\\
&\vphantom{.}&\,\,\,\,=  {1\over b_1}\sum_{\lambda,\omega\in H^1(X_2, \ZZ_2) } Z_{\cT}[\lambda] (-1)^{\mathrm{Arf}(\omega)+ \int_{X_2} \omega \cup \Lambda  + \int_{X_2}\lambda \cup \omega }~.\no
\eea
Using \eqref{2danom},  the above partition function is equivalent to 
\bea
 {1\over b_1}\sum_{\lambda,\omega \in H^1(X_2, \ZZ_2) } Z_{\cT}[\lambda]\, (-1)^{  \mathrm{Arf}(\lambda+\omega)+\int_{X_2} \omega \cup \Lambda }~,
\eea
where we have used 
\bea
\mathrm{Arf}(A+B) + \mathrm{Arf}(A)+ \mathrm{Arf}(B) = \int A\cup B~.
\eea
Finally, changing variables from $\omega$ to $\omega + \lambda$ gives 
\bea
{1\over \sqrt{b_1}} \sum_{\lambda \in H^1(X_2, \ZZ_2)}Z_{\cT}[\lambda]\, (-1)^{  \mathrm{Arf}(\Lambda)+\int_{X_2} \lambda \cup \Lambda }
\eea
upon noting that $\sum_\omega (-1)^{ \mathrm{Arf}(\omega)} = \sqrt{b_1}$. This is indeed equal to the original theory $Z_{\cT/\ZZ_2^{(0)}}[\Lambda]$ up to a factor of $(-1)^{ \mathrm{Arf}(\Lambda)}$, which can be removed by a $T$ transformation. 
To summarize, we have shown that the theory $\cT/\Z_2^{(0)}$ is self-dual under $TST$.

\subsection{Making contact with Kramers-Wannier duality}

Note that the theory $\cT/\Z_2^{(0)}$ discussed above is not the usual bosonization map. Hence if $\cT$ is a free massless Majorana fermion,  the theory $\cT/\Z_2^{(0)}$ whose self-duality we just discussed is not the critical Ising model. In particular, $\cT/\Z_2^{(0)}$ depends on the choice of spin structure: $Z_{\cT/\Z_2^{(0)}}[\Omega]$ is not invariant under the change of spin structure $\Omega\to \Omega+\Lambda$. 

Instead, the bosonic theory $\cB$ is given by \cite{Ji:2019ugf, Karch:2019lnn}
\begin{equation}
Z_{\cB}[\Omega] = {1\over \sqrt{b_1}}\sum_{\lambda \in H^1(X_2, \ZZ_2)} Z_{\cT}[\lambda] (-1)^{\int_{X_2} \lambda \cup \Omega+ \mathrm{Arf}(\Omega)}
\end{equation}
which is related to $\cT/\Z_2^{(0)}$ via
\begin{eqnarray}
\cB=T \left( \cT/\Z_2^{(0)}\right)~.
\end{eqnarray}
Using the self duality of $\cT/\Z_2^{(0)}$ obtained above, i.e. $\cT/\Z_2^{(0)}= TST(\cT/\Z_2^{(0)})$, we find that
\begin{eqnarray}
S(\cB)= ST(\cT/\Z_2^{(0)})= T(\cT/\Z_2^{(0)}) = \cB~.
\end{eqnarray}
This is exactly the standard Kramers-Wannier duality when $\cB$ is the critical Ising model.

\end{document}